# Securing the supply of graphite for batteries

K. Bhuwalka[1,2], H. Ramachandran[3], S. Narasimhan[2,3], A.Yao[2,3], J. Frohmann[4], L. Peiseler[1], W. C. Chueh[2,3,4], A. Boies[5], S. J. Davis[6,1], and S. M. Benson[4]

[1] Precourt Institute for Energy, Stanford University, Stanford, CA 94305, USA
[2] Applied Energy Division, SLAC National Accelerator Laboratory, Menlo Park, CA 94305, USA
[3] Department of Materials Science and Engineering, Stanford University, Stanford, CA 94305, USA
[4] Department of Energy Science and Engineering, Stanford University, Stanford, CA 94305, USA
[5] Department of Mechanical Engineering, Stanford University, Stanford, CA 94305, USA
[6] Department of Earth System Science, Stanford University, Stanford, CA 94305, USA

## Abstract

**Surging demand for graphite in energy storage applications has led to concerns about supply chain security for manufacturers and nations globally. Currently, China produces over 92% of graphite for anodes, posing a risk for industries reliant on graphite supply. Here, we systematically assess the costs of producing natural and synthetic battery-grade graphite in the U.S. and China using process-based cost models. We find that production costs in the U.S. are higher than those in China by 100-200%, so scaling production in the short-term will require significant policy support. We use our models to explore opportunities to improve the competitiveness of graphite production outside China, finding that lower financing rates and improved shaping yields can together reduce costs by 25-30%. Further implementing other cost reduction strategies, such as improving process throughput or lowering equipment costs, can achieve 35-40% lower costs. Finally, we discuss how innovative graphite production processes like methane pyrolysis and catalytic graphitization may provide competitive pathways.**

## Introduction

Global demand for graphite used in battery anodes is surging as consumption of batteries in electric vehicles (EVs), grid-scale energy storage systems (ESS), and consumer electronics grows. From 2018 to 2023, graphite demand grew by over 50% to 3.6 million metric tons (Mt), with most of the increase (89%) attributed to battery demand.[1] Demand specifically for use in batteries is projected to continue growing rapidly and reach 2.9 Mt in 2028[1] (**Fig 1a**) —an 1,800% increase in a decade.

Currently, China produces over 92% of graphite used for battery anodes (**Fig 1b**), posing a significant supply risk to consumers and nations across the world.[2] Graphite exports have already become a point of geopolitical leverage: in late 2024, China restricted graphite exports in response to U.S. limits on semiconductor exports.[3] Although diversification of graphite supply chains could reduce trade risks, the low prices of Chinese graphite presents a major challenge to scaling up production elsewhere. In response, many countries are incentivizing local production of graphite.[4] In 2025, the U.S. raised tariffs on graphite imports from China to 160%[5]. Although such aggressive trade policies could boost the competitiveness of producers outside China, higher prices of graphite will increase costs of EVs and ESS, thereby slowing the transition to clean energy systems.[6,7]

Given the urgency of scaling battery-grade graphite production globally, research and investment should prioritize processes that are most likely to be competitive. Although there have been some technical studies of new graphite production technologies and life-cycle assessments of conventional production routes,[8–15] there is no comprehensive assessment of current graphite production processes and drivers of cost. Broadly speaking, battery graphite can be produced from fossil fuel feedstocks (synthetic graphite) or geological ore deposits (natural graphite) (**Fig 1c**). As researchers and entrepreneurs begin developing new production methods[16,17], a bottom-up understanding of production costs for both these routes is critical for benchmarking and assessing the viability of new technologies. Moreover, quantifying the key cost components is essential to guide innovation and research towards process improvements that will have the largest impact on improving project economics and scaling supply.

Here, we use industry-vetted data on key input parameters, including cost of raw materials, energy, labor, equipment, and construction to develop detailed process-based cost models for natural and synthetic pathways to produce battery-grade graphite. We define a baseline value and a range for parameters relating to process operations, input costs, and capital financing which our model uses to calculate the overall costs of production. To ensure the assessment represents state-of-the-art production processes, model results and parameters were validated via written comments from 15 companies in the graphite value chain, as well as feedback during two workshops attended by 120 experts from 70 organizations (Details in SI Section I) The models support “apples-to-apples” comparison of graphite production costs across different processing routes and geographic regions. Further, we run a Monte-Carlo simulation of graphite production costs across a range of inputs to systematically investigate the sensitivity of costs to parameters such as feedstock costs, process efficiencies, and capital financing rates, using the results to set out quantitative targets for economic viability and highlight innovation and policy priorities. Details of our analytic approach are described in Methods.

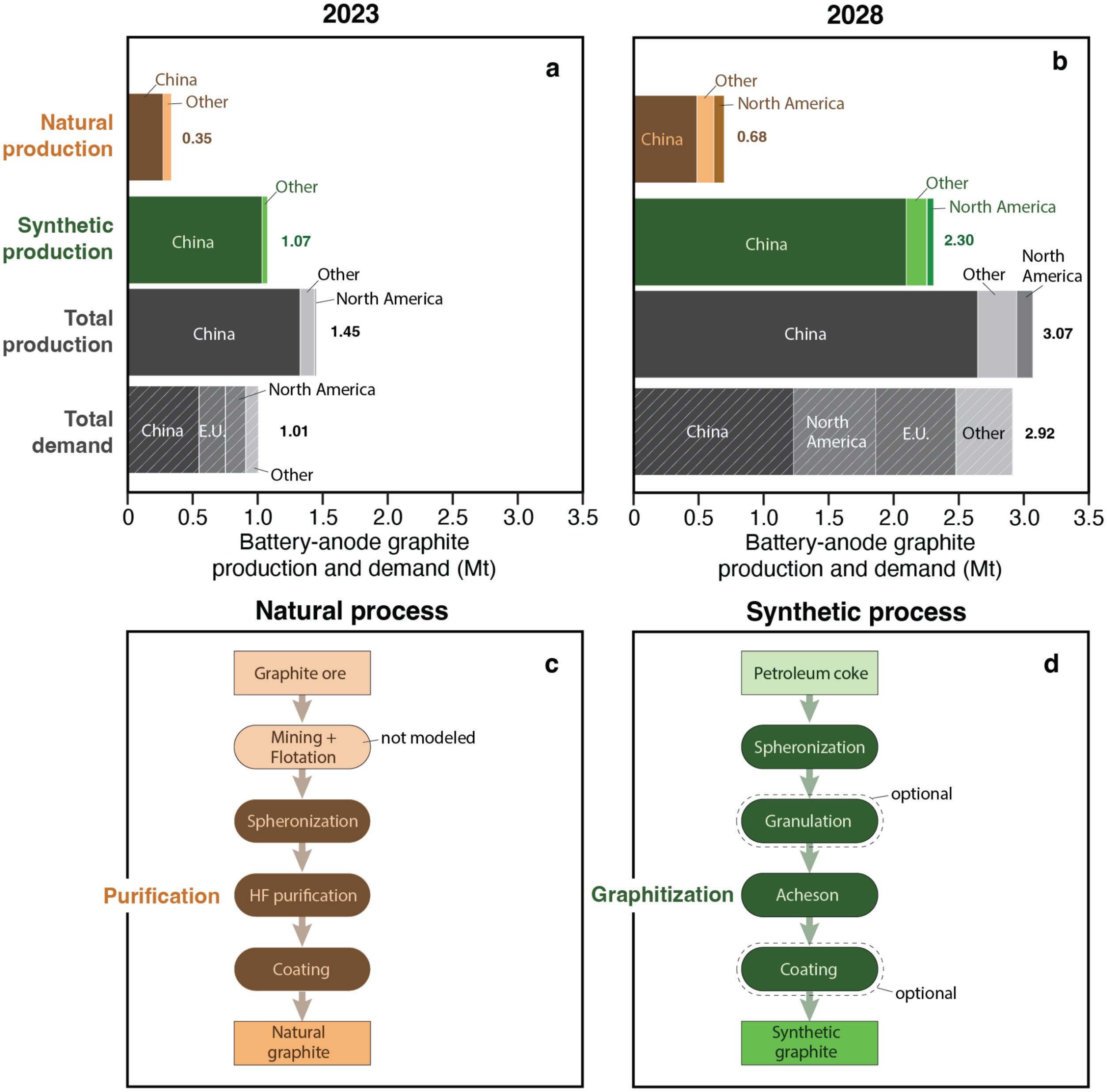

**Figure 1 | Overview of the battery-grade graphite market and production processes.** Battery-anode graphite production and demand in 2023 (**a**) and projections for 2028 (**b**). Demand (hatched bars) and production (solid bars) are segmented by region and material type: natural or synthetic. Demand and Supply data is from Oxford Economics[1] who cite Benchmark Mineral Intelligence as the primary source. Wire diagrams show natural (**c**) and synthetic (**d**) production processes. Processes with dashed outlines are optional processes that may or may not be used depending on the quality requirements of the end consumer.

**Processes of graphite production**

Graphite used in the negative electrode (anode) of a lithium-ion battery has much more stringent quality requirements than graphite used in electric-arc furnace electrodes or in lubricants because quality strongly affects the energy density, cycle life, and charge rates of batteries (See SI.A for details). Such quality requirements necessitate highly controlled production processes for battery-grade graphite.

There are two main pathways for producing battery-grade graphite today: one based on mined flake graphite ("natural graphite") and the other on petrochemical feedstocks ("synthetic graphite") (**Fig 1**). Historically, natural graphite has been cheaper but recent rapid declines in prices for synthetic graphite have caused their market share to reach 75% in 2025[2]**.** Synthetic graphite is produced from specific forms of petroleum coke, predominantly needle coke. The coke is spheronized into 10-20 micron sized particles ("shaping"), and then primary particles are agglomerated into secondary particles by carbonizing with a carbonaceous pitch ("granulation"). Following this, the coke particles are graphitized in an Acheson furnace by heating to temperatures of around 3000 °C via strong electrical currents and then cooling slowly over 18-20 days. Within the furnace, the coke particles are contained in cylindrical graphite crucibles and surrounded by a large amount of coke "packing material" which isolates them from air. In contrast, natural graphite production starts with mining flake graphite ore, which is in turn concentrated and segregated based on size, shaped into spheres, and purified using a combination of heating and acid or alkali treatment (typically, HF purification; Fig. 1c). If acids are used, they are neutralized with lime. Both natural and synthetic graphite are coated in an inert nitrogen environment with a pitch as a final step.

Production costs are a combination of the cost of equipment, installation, and facility construction (together termed capital costs) as well as the costs of utilities, labour, reagents like acids, consumables like crucibles, and feedstock materials (together termed operating costs). These individual processes are described further in the Methods section, and all the cost factors are listed in SI Tables S3-S13.

**Graphite production costs in the U.S. and China**

Our modeling shows that production costs per metric ton ($/t) of both natural and synthetic battery-grade graphite are substantially higher in the U.S. than in China (**Fig. 2)**. Operating costs to produce synthetic graphite in the U.S., which are dominated by the costs of electricity and consumables, are substantially higher than in China (y-axis in Fig. 2b), consistent with industry reports[18–20], Interestingly, the most significant difference in costs between the regions is the higher capital costs of U.S. plants compared to plants in China (x-axis in Fig 2b). The capital costs include the cost of equipment, installation, construction and contingency and average $28,300/tpa and $22,000/tpa for synthetic and natural graphite, respectively, in the U.S. compared to $9,000/tpa and $5,600/tpa for synthetic and natural graphite, respectively, in China.

Diagonal contour lines in **Figure 2** represent the overall production costs per ton of graphite for a project with a cost structure represented by the two axes. For a project to be competitive, the price of graphite must be lower than the overall production cost represented by the contour line. Prices for both natural and synthetic graphite produced in China have fallen dramatically between 2022 and 2025 (dark grey bands in Fig. 2). While 55% and 50% of modeled costs of U.S.-produced natural and synthetic graphite, respectively, would have been competitive with China's 2022 prices, none are competitive if prices remain at 2025 levels.

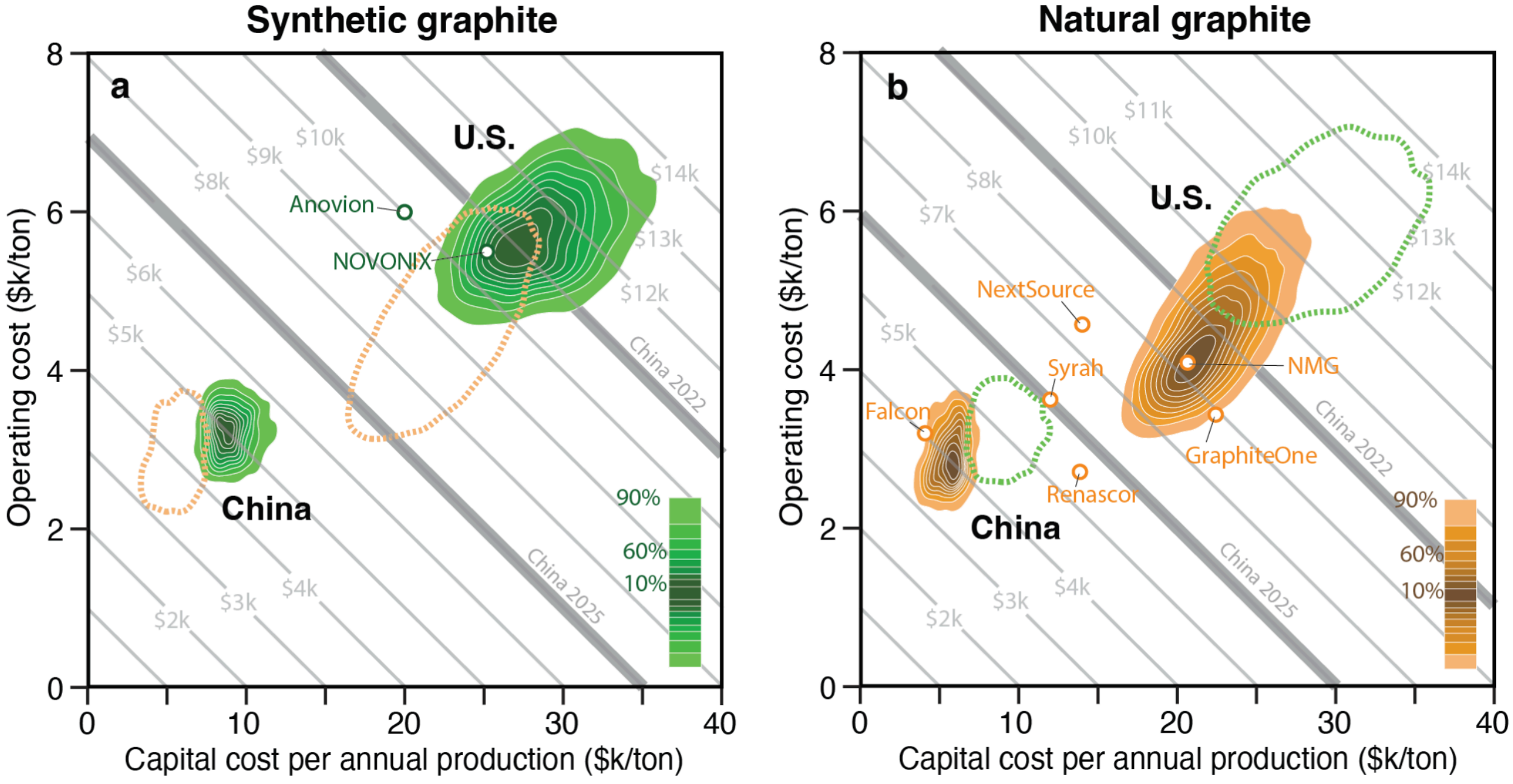

**Figure 2 | Modeled costs of graphite production.** Production of natural (**a**) and synthetic (**b**) graphite in thousands of dollars per metric ton are shown as a function of operating costs (y-axes) and capital costs per tonne of annual production (x-axes). The distributions represent the range of costs resulting from the Monte-Carlo simulations with the darkness of shading indicating the share of modeled results falling within the area (90% of simulations fall within the entire colored region). The grey diagonals represent the overall production costs for a project with a given operating and capital costs (capital costs are annualized over 10 years with a 15% cost of capital). The bold diagonals represent historic average prices of high-quality battery-grade graphite from China in 2022 and 2025[2,21]. Circles show publicly reported costs from indicated companies outside of China (in some cases adjusted for consistency; see Table S2 for more details).

**Drivers of synthetic graphite production costs**

Our analysis reveals a striking disparity between production costs of synthetic graphite in the U.S. and China, with U.S. costs more than double ($12,300/t; **Figure 3a**) those of China ($5,400/t; **Figure 3b**) in the "baseline" case (i.e. when input parameters are set to their baseline value according to Tables S3-S13). This difference is due to both higher operating and capital costs in the U.S. ($6,000/t compared to $3,800/t in China and $6,200/t compared to $1,600/t in China, respectively; Figs. 3a and 3b).

**Figures 3a and 3b** show which processes contribute most to synthetic graphite production costs in the U.S. and China, respectively. The largest portion of costs in both regions is related to the graphitization process, which involves construction of large Acheson furnaces and substantial use of electricity and other consumables (graphitization costs $6,300/t in the U.S. and $2,400/t in China, or 51% and 44% of total costs, respectively). **Figure 3c** further compares different costs between the two regions. Across each category, capital costs are much lower in China ($430-$770/t) than the U.S. ($1600-$2500/t).

Petroleum coke feedstock represents a larger cost in the U.S. (~$1200/t) compared to China (~$800/t) despite the U.S. having domestic petroleum coke supply because producers in China have learned to substitute some high-grade needle coke feedstocks with lower-grade coke without compromising on product quality[22]. Electricity prices, a crucial factor in the electricity-intensive Acheson process, are also lower in inner Mongolia–where most graphitization in China occurs– than in the southeast U.S. (difference in costs of $240/t).

Finally, consumables such as "packing coke" in furnaces are a large driver of costs because two tons of packing coke are consumed per ton of synthetic graphite produced. In the U.S., purchasing and disposing of packing coke accounts for ~$1000/t (8%) of synthetic graphite cost, whereas in China up to 80% of this packing coke is recycled for use in other industries such as steelmaking[22]. The graphite crucibles that contain the coke during graphitization are also a substantial cost driver since they can only be reused 4-5 times before deteriorating. Moreover, there is no existing supply chain for crucibles in the U.S., leading to supply chain vulnerability and costs from shipping.

**Drivers of natural graphite production costs**

There are similar differences in costs to produce natural graphite in the two regions (baseline cost is $8,400/t in the U.S. and $3,600/t in China; **Figures 3d and 3e**. The difference is again driven by capital costs ($4,100/t in the U.S. and $700/t in China, or 49% and 20% of total costs, respectively), but–unlike for synthetic graphite–no single process step dominates.

**Figure 3f** further discretizes the cost differences, showing the importance of cheaper equipment, installation, and construction costs in China. The feedstock graphite concentrate also represents a sizable proportion of natural graphite production costs in the U.S. ($1,600/t or 19%).

Compared to SG, energy and consumables costs make up a smaller portion of costs of natural graphite production since it does not require high-temperature graphitization. However, there are some additional costs for the acids used in the purification step ($400/t). In regions where there is resistance to use of hydrofluoric acid due to health hazards, alternative pathways such as carbochlorination, high-temperature purification, and acid-alkali roasting can be used, but lead to higher costs and lower purity than acid purification[23] (discussed further in SI.F).

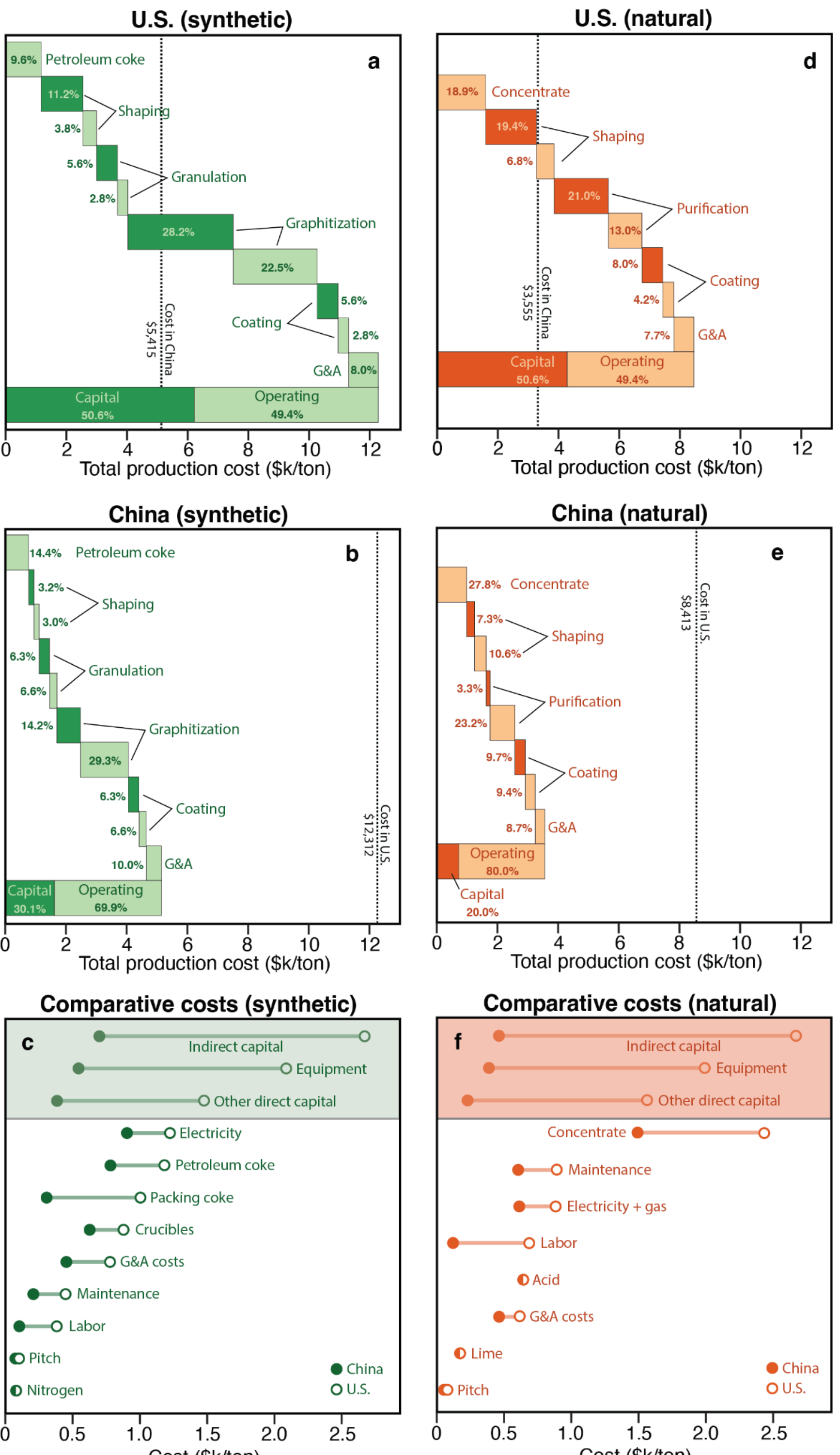

**Figure 3 | Baseline costs of graphite production, segmented by type.** Costs of producing synthetic graphite for batteries is much higher in the U.S. (**a**) than in China (**b**) largely due to capital costs, especially for graphitization. Cost differences in the two countries are similar for production of natural graphite (**d, e**). All bars show "baseline" costs, which are the costs when input parameters are set to their baseline value described in Tables S3-13. Capital and operating costs differentiated by dark and light shading, respectively; Panels **c** and **f** further decompose relevant costs in the two countries. G&A refers to general and administrative costs.

## Increasing the competitiveness of graphite production

**Figure 4** demonstrates a potential pathway to greatly reduce U.S. graphite production costs. From top to bottom, the bars show the cumulative cost reductions of various interventions, ordered in decreasing order of interventions that have the greatest impact (importance calculations in SI.H). Across both natural and synthetic graphite production, the most impactful way to lower costs is to lower the cost of capital, which is the required rate of return a company must earn on its investments to satisfy its investors and lenders. Reducing the cost of capital from the baseline 15% to 5% reduces the cost of synthetic and natural graphite by 18% (from $12,300/t to $10,100/t for synthetic, and from $8,400/t to $7,000/t for natural). Financing rates depend on project risk, and given low graphite prices and resulting competitiveness challenges, private investors are unlikely to finance a project if expected rates of return are lower than 15%. The cost of capital could be lowered if a producer signs an offtake agreement with a consumer, which provides certainty on future revenue. Even so, private investors are unlikely to be satisfied with a 5% return, but such low costs of capital can be achieved via government financing for strategic industries.

The second most impactful lever is the yield of the shaping process. For synthetic graphite, boosting shaping yields from 70% to 90% can further cut costs by 6% (from $10,100/t to $9,500). Natural graphite, with typically lower yields, sees an even greater impact: increasing yields from 50% to 65% reduces costs by 11% (from $7,000/t to $6,200/t). As projects scale and optimize their processes over time, yields will likely increase, thereby reducing both feedstock waste and equipment needs for the same output quantity.

The third and fourth most important interventions for natural graphite are also shaping-related. A 20% increase in throughput (tonnes of material processed per hour) for shaping, which corresponds to a 20% decrease in process capital costs per tonne of output, can reduce costs further from $6,200/t to $5,800/t (6.5% lower). Moreover, if the low shaping yield and higher throughput is combined with low graphite concentrate prices ($600/t), costs decrease by another 5% ($5,800/t to $5,500/t). While concentrate prices from China are currently lower than $600/t, it is projected that the concentrate market will soon enter a supply deficit, likely raising prices.[24] Finally, for natural graphite, innovation that increases the throughput of the purification step by 20% drives down total costs from $5,500 to $5,300 (4% lower).

For synthetic graphite, the third most important intervention is increasing the throughput of graphitization. A 20% increase in throughput is akin to holding the throughput constant but reducing graphitization capital expenditures by 20%. Such an intervention can reduce costs by 4% from $9,500/t to $9,100/t. Because of the high energy consumption via Acheson furnaces, the 4th most important parameter is the price of electricity. Sourcing electricity at 4¢/kWh rather than the baseline 6.5 ¢/kWh would further reduce costs from $9,100/t to $8,600/t (5.5% lower). While strategic selection of production location can help, such low electricity prices can be hard to access: In 2024, the lowest average industrial electricity price of any state was 5.5c/kwh (New Mexico)[25]. Although we model the most commonly-used Acheson furnaces in this paper, some producers have started using new furnace technologies such as continuous or box furnaces, which promise both higher effective throughput and lower electricity consumption. Such furnaces may lead to potential cost savings, but have technical challenges in producing graphite of the desired quality (trade-offs discussed in SI.E).

Finally, reducing the cost of consumables via circular processes is another lever that can drive down synthetic graphite costs. Used (and partially-graphitized) packing coke may be sold to the steel industry, or be recycled to make other graphite products. If 80% of the packing coke is recycled for use in other industries without any loss in value (net consumption reducing from 2t to 0.4t per tonne of product), costs reduce further $8,600/t to $7,700/t.

If producers in the U.S. could implement all the cost reduction strategies described here successfully, costs would be lower by 37% for both synthetic and natural graphite. Another potential strategy to reduce capital costs in the U.S. is to import less-expensive equipment from China, which alone can lead to a cost reduction of 15%-20% compared to the baseline (SI.D). A 40-50% reduction in costs can help bring production costs to levels that are competitive with prices, but many of the process innovations require learning-by-doing which is unlikely until an industry is stood up. Consequently, building supply of graphite in the West requires either i) a sustained price premium which would make projects competitive or ii) disruptive innovation that reduces costs significantly.

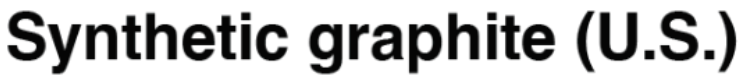

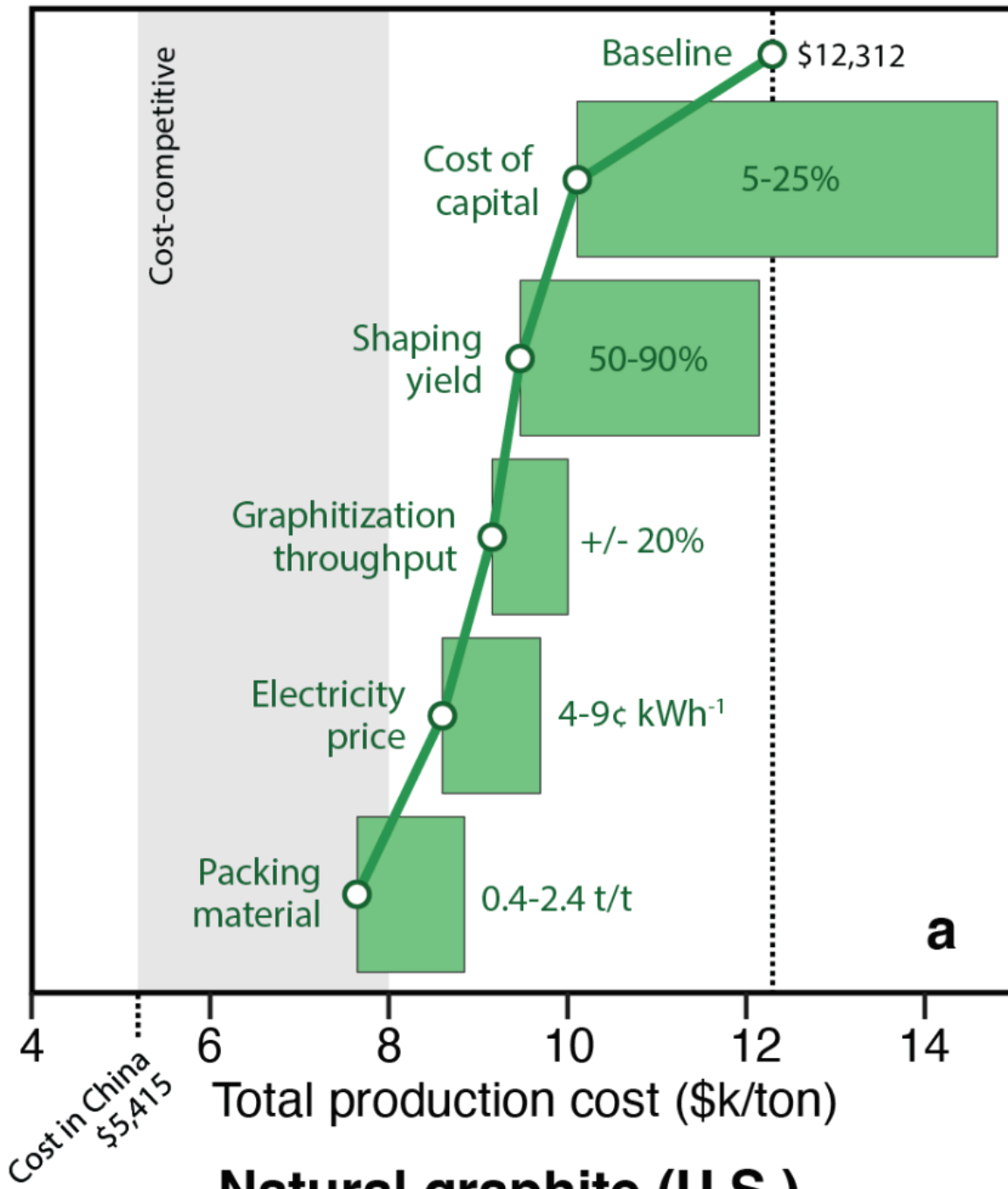

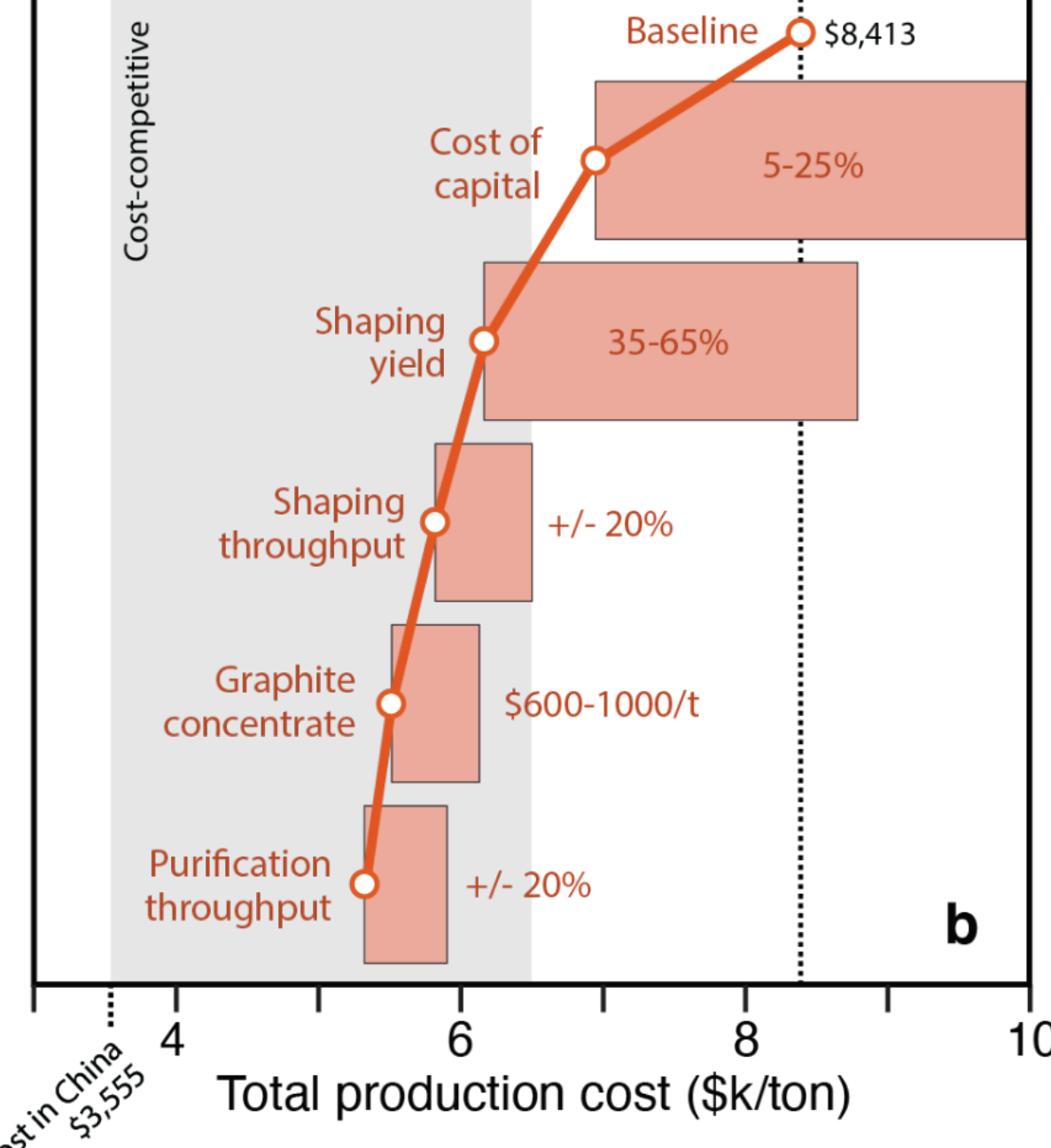

**Figure 4 | Pathways for cost reductions in the U.S.** Bars show the cumulative effect of changing input parameter values on the total production costs of U.S. synthetic (**a**) and natural (**b**) graphite production. The parameter ranges shown may be feasible, based on literature and interviews. Parameters are changed one-by-one from top to bottom, so the plots highlight a single pathway to cost reduction from baseline costs at the top and assuming maximum improvements from each parameter. Interventions in decreasing order of importance, measured based on SHAP values (SI Section H). The grey regions are the range of prices for high-grade synthetic and natural graphite in China in 2025[17], and represent costs at which U.S. produced graphite could be sold competitively.

## Discussion

The concentration of global graphite production in China presents a significant supply chain security risk, a concern already underscored by recent geopolitical actions such as export restrictions and tariffs. Graphite's importance extends beyond lithium-ion batteries to other strategic sectors like nuclear power, amplifying the need for supply diversification. However, this diversification comes at a cost that must be borne by someone—either directly by consumers through higher battery prices or indirectly through government policies and taxes. For example, an EV would be ~$700 more expensive to produce in the U.S. if synthetic graphite prices were based on our modeled U.S. costs instead of China costs, which would in turn hurt sales of domestic manufacturers and extend fossil fuel emissions from transportation [7]. This study provides a quantitative framework for this ongoing political and industrial debate about the costs and benefits of investing in supply diversification for graphite.

We find that production costs in the U.S. are over 2 times higher than in China, rendering production economically unviable without significant cost reductions and policy support. Since capital costs are the most dominant driver of the cost of graphite, the most effective policy levers are those that either provide a sustained price premium or offer low-cost loans. Offtake agreements by consumers can reduce project risk and unlock more favorable private financing. Ultimately, near-term feasibility depends on both automotive companies and policymakers' willingness to pay premiums for graphite from producers outside China.

However, price premiums are not sustainable in the long-run as the higher costs are borne by either consumers or taxpayers. Breakthrough innovations are required that can leapfrog cost challenges, and our paper highlights key priority areas for innovation, as well as cost targets for new technologies to be viable. For researchers focusing on improving a particular step in the graphite production process, the paper quantitatively benchmarks the costs of necessary pre - and post- processing steps and allows them to evaluate the overall feasibility of their technology.

Three promising pathways for future R&D are (1) methane pyrolysis, (2) catalytic graphitization of bio-char, and (3) end-of-life battery recycling. Methane pyrolysis leverages large natural gas reserves in the U.S. and elsewhere to produce hydrogen and carbon. Previous work has focused on optimizing hydrogen yields and producing lower-value amorphous carbon as by-products, with techno-economic assessments showing it is possible to produce amorphous carbon at less than $1000/t [26–28]. However, since graphitization is the primary cost driver in battery-grade graphite production, pyrolysis will only be viable if we improve on methods which directly produce crystalline graphite with minimal impurities. [29,30] If graphitic carbon that meets battery-grade specifications can be produced for $1000/t, the shaping and coating steps would take total costs to ~$4000/t in the U.S. and the product would be competitive.

Catalytic graphitization of biochar presents another alternative that utilizes agricultural waste. Use of a catalyst, often iron powder, facilitates the rearrangement of carbon atoms into the layered structure of graphite and reduces graphitization temperatures from 3000°C to around 1200°C.[16,31,32] While the technology is promising, future research should focus on reducing the need for intensive purification that is currently required to remove the catalyst and reach 99.95% product purity. The purification and coating steps in the U.S. cost $4000/t so catalytic graphitization must cost less than $1000/t to bring prices to China levels. Finally, recycling of graphite from end-of-life batteries could provide a secure source of supply. Graphite recycling has potential because it already meets specifications and does not need to undergo graphitization, granulation and shaping which represents over 75% of synthetic graphite production costs. However, processes need to overcome challenges in economically removing impurities and restoring the graphite's original structure.

Even if new technologies were feasible today, testing and qualification of new material could take up to five years, including lab-bench, pilot, and full operation testing to assess the performance of the material in a battery under different conditions.[33] If material takes years to qualify, the delayed revenue can make project economics unsustainable. Industry coordination is needed to reduce qualification times and help producers get to market faster: one method could be to identify test metrics that can predict performance implications of switching to the new materials.

Our study underscores the economic challenges facing the Western graphite industry while identifying key leverage points for policymakers and innovators. Short-run price premiums and financing support for graphite projects can bridge the immediate gap, while significant R&D investment into new production routes and a coordinated effort to streamline product qualification will be crucial for long-term competitiveness.

## Methods

In this paper, we build process-based cost models to estimate the cost of producing graphite for battery anodes. To build the models, we identify the main inputs and outputs of the production process and assign costs to each process step. Key input parameters include the cost of raw materials, energy, labor, equipment, and construction. Data on input parameters is sourced from feasibility studies and verified via extensive feedback from industry. Industry feedback was sourced via individual interviews, written responses to draft manuscripts, and two workshops that brought together over 150 experts in the graphite value chain.

For a desired plant capacity, we scale the capital investment and estimate the number of production lines needed for each process. We annualize the capital costs based on a rate representing the weighted average cost of capital (WACC) and derive a total cost per ton of graphite produced (Details in SI Section G). The model is applied to compare costs of AAM produced in the U.S. and China.

We assume three major process steps for the production of natural graphite. The 100-mesh portion (< 150 um) of graphite concentrate is micronized and spheronized in order to reduce the particle size and to turn the graphite flakes into spherical particles with a 50th percentile size of 20 micron and tapped density between 0.85 and 0.95 g/cc (NMG; Table 13-13). Second, graphite is purified to a concentration of >99.95% via acid leaching. Finally, the purified spherical graphite is coated with a thin layer of petroleum pitch to reduce the surface area of graphite and improve both the first cycle efficiency and rate performance of the material. The main inputs for each process step are presented in the Supporting Information (SI).

For battery-grade synthetic graphite production, calcined needle coke is first spheronized to produce primary particles. Primary particles are then agglomerated into larger secondary particles (10-20 micron) by heating them with petroleum pitch ('Granulation process'). Graphitization can take many forms, but the Acheson furnace process dominates production today. Here, spherical coke particles are contained in cylindrical graphite crucibles surrounded by a large amount of coke "packing material" which isolates them from air. The furnaces are powered by strong currents passing through high-power DC electrodes, and the crucibles are resistively heated to around 3,000ºC. After several hours of sustained passage of current, the furnace is allowed to cool slowly over the next 18-20 days before the material is removed. Finally, the graphitized particles can be optionally pitch coated to reduce surface area and improve performance, in a process similar to the natural graphite route. We assume that the resulting graphite has a specific capacity greater than 350mAh/g.

For all the processing pathways, we run the cost model over a range of input parameters by conducting a Monte-Carlo simulation over uniformly distributed input parameters. We discuss the distribution of the simulated costs in Figure 2, and use the results of the simulations to identify which parameters have the most significant impact in reducing production costs in Figure 4. To identify which input parameters have the maximum impact on production costs, we conduct SHAPley sensitivity analysis (described in SI Section H)

## References

1. Oxford Economics. *Enabling North American Graphite Growth*. (2024).
2. Benchmark Mineral Intelligence. *Benchmark Mineral Intelligence* https://www.benchmarkminerals.com/.
3. Benson, E. & Denamiel, T. China's New Graphite Restrictions. (2023).
4. Battery-makers, graphite producers clash over tariffs on Chinese supply. *S&P Global Market Intelligence* https://www.spglobal.com/marketintelligence/en/news-insights/latest-news-headlines/battery-makers-graphite-producers-clash-over-tariffs-on-chinese-supply-81016370.
5. US graphite miners ask Washington to impose 920% tariff on Chinese rivals | Reuters. https://www.reuters.com/markets/commodities/us-graphite-miners-ask-washington-impose-920-tariff-chinese-rivals-2024-12-18/.
6. Allcott, H., Kane, R., Maydanchik, M. S., Shapiro, J. S. & Tintelnot, F. The Effects of ``Buy American'': Electric Vehicles and the Inflation Reduction Act.
7. Wang, H. *et al.* China's electric vehicle and climate ambitions jeopardized by surging critical material prices. *Nat Commun* **14**, 1246 (2023).
8. Surovtseva, D., Crossin, E., Pell, R. & Stamford, L. Toward a life cycle inventory for graphite production. *Journal of Industrial Ecology* **26**, 964–979 (2022).
9. Engels, P. *et al.* Life cycle assessment of natural graphite production for lithium-ion battery anodes based on industrial primary data. *Journal of Cleaner Production* **336**, 130474 (2022).
10. Dai, Q., Kelly, J. C., Gaines, L. & Wang, M. Life Cycle Analysis of Lithium-Ion Batteries for Automotive Applications. *Batteries* **5**, 48 (2019).
11. Whattoff, P. *Prospective Life Cycle Assessment of Epsilon Carbon Ltd Synthetic Graphite Anode Manufacturing*. https://www.epsilonam.com/pdf/life-cycle-report.pdf#page=67.11.

12. Salas, S. D. & Dunn, J. B. Methane-to-graphite: A pathway to reduce greenhouse gas emissions in the U.S. energy transition. *Resources, Conservation and Recycling* **210**, 107832 (2024).
13. Trotta, F. *et al.* A Comparative Techno-Economic and Lifecycle Analysis of Biomass-Derived Anode Materials for Lithium- and Sodium-Ion Batteries. *Advanced Sustainable Systems* **6**, 2200047 (2022).
14. Gao, S. W., Gong, X. Z., Liu, Y. & Zhang, Q. Q. Energy Consumption and Carbon Emission Analysis of Natural Graphite Anode Material for Lithium Batteries. *Materials Science Forum* **913**, 985–990 (2018).
15. Lee, S.-M., Kang, D.-S. & Roh, J.-S. Bulk graphite: materials and manufacturing process. *Carbon letters* **16**, 135–146 (2015).
16. You, H. *et al.* Sustainable Production of Biomass‐Derived Graphite and Graphene Conductive Inks from Biochar. *Small* **20**, 2406669 (2024).
17. Dey, S. C. *et al.* Catalytic graphitization of pyrolysis oil for anode application in lithium-ion batteries. *Green Chem.* **26**, 8840–8853 (2024).
18. Allaire, A., Eng, P., Martel, B.-O. & Geo, P. *Nouveau Monde Graphite 2025 Feasibility Study*. (2025).
19. de Wit, D. R., Siegert, J., Audet, M.-A. & Moryoussef, P. Falcon Energy Materials Natural Graphite Technical Report.
20. Barr Engineering Co. Graphite Creek Project NI 43-1010 Technical Report and Feasibility Study. *Technical Report* (2025).
21. Shanghai Metals Market. https://www.metal.com/.
22. Wang, F. *et al.* Carbon Footprint and Decarbonization Potential of Battery-Grade Synthetic Graphite. *ACS Sustainable Chem. Eng.* **13**, 8298–8306 (2025).
23. Huang, S., Wang, W., Cui, Q., Song, W.-L. & Jiao, S. Assessment of Spherical Graphite for Lithium-Ion Batteries: Techniques, China's Status, Production Market, and Recommended

Policies for Sustainable Development. *Advanced Sustainable Systems* **6**, 2200243 (2022).

24. ESG of graphite: how do synthetic graphite and natural graphite compare? *Benchmark* https://source.benchmarkminerals.com/article/esg-of-graphite-how-do-synthetic-graphite-and-natural-graphite-compare (2022).
25. Electric Power Monthly - U.S. Energy Information Administration (EIA). https://www.eia.gov/electricity/monthly/epm_table_grapher.php.
26. Shokrollahi, M., Teymouri, N., Ashrafi, O., Navarri, P. & Khojasteh-Salkuyeh, Y. Methane pyrolysis as a potential game changer for hydrogen economy: Techno-economic assessment and GHG emissions. *International Journal of Hydrogen Energy* **66**, 337–353 (2024).
27. Okeke, I. J., Saville, B. A. & MacLean, H. L. Low carbon hydrogen production in Canada via natural gas pyrolysis. *International Journal of Hydrogen Energy* **48**, 12581–12599 (2023).
28. Moghaddam, A. L. *et al.* Methane pyrolysis for hydrogen production: navigating the path to a net zero future. *Energy Environ. Sci.* **18**, 2747–2790 (2025).
29. Dermühl, S. & Riedel, U. A comparison of the most promising low-carbon hydrogen production technologies. *Fuel* **340**, 127478 (2023).
30. Cornejo, A. & Chua, H. T. Process for producing hydrogen and graphitic carbon from hydrocarbons. (2022).
31. Shi, Z. *et al.* Bio-based anode material production for lithium–ion batteries through catalytic graphitization of biochar: the deployment of hybrid catalysts. *Sci Rep* **14**, 3966 (2024).
32. Lower, L. *et al.* Catalytic Graphitization of Biocarbon for Lithium‐Ion Anodes: A Minireview. *ChemSusChem* **16**, e202300729 (2023).
33. Associates, B. Lithium-ion Battery Recycling: Scale up and Battery Qualification. *Battery Associates* https://www.battery.associates/post/lithium-ion-battery-recycling-scale-up-and-battery-qualification (2022)